\documentclass[12pt, letterpaper]{article}
\usepackage[top=1in, bottom=1.5in, left=1in, right=1in]{geometry}
\usepackage[utf8]{inputenc}
\usepackage{booktabs}
\usepackage{hyperref}

\title{\Large Characterizing Variable Stars in a Single Night with LSST}
\author{Eric D. Feigelson with Frederica Bianco and Sara Bonito}

\begin{document}

\maketitle

\begin{abstract}
Stars exhibit a bewildering variety of variable behaviors ranging from explosive magnetic flares to stochastically changing accretion to periodic pulsations or rotations.  The principal LSST surveys will have cadences too sparse and irregular to capture most of these phenomena.  A novel idea is proposed here to observe a single Galactic field, rich in unobscured stars, in a continuous sequence of $\sim 15$ second exposures for one long winter night in a single photometric band.  The result will be a unique dataset of $\sim 1$ million regularly spaced stellar lightcurves.  The lightcurves will gives a particularly comprehensive collection of dM star variability.   A powerful array of statistical procedures can be applied to the ensemble of lightcurves from the long-standing fields of time series analysis, signal processing and econometrics.  Dozens of `features' describing the variability can be extracted and subject to machine learning classification, giving a unique authoritative objective classification of rapidly variable stars.  The most effective features can then inform the wider LSST community on the best approaches to variable star identification and classification from the sparse, irregular cadences that dominate the LSST project.  
\end{abstract}

{\small
\section{White Paper Information}
Professor Eric D. Feigelson \\
Departments of Astronomy \& Astrophysics and of Statistics \\
Penn State University, University Park PA 16801  \\
Email: edf@astro.psu.edu \\

\noindent Co-signers:\\
Frederica Bianco, Co-chair, LSST TVS Science Collaboration (fb55 at nyu.edu) \\
Sara Bonito, Spokesperson,  LSST TVS Task Force on Deep-drilling fields and mini-surveys (rosaria.bonito@inaf.it)

\begin{enumerate} 
\item {\bf Science Category:} Exploring the Changing Sky
\item {\bf Survey Type Category:} Other (single night continuous observation) 
\item {\bf Observing Strategy Category:} an integrated program with science that hinges on the combination of pointing and detailed observing strategy 
\end{enumerate}  
}
\clearpage

\section{Scientific Motivation}
{\footnotesize

Astronomers have become used to poorly cadenced data in the study of variable stars.  Irregular observations with wide gaps due to solar motion and erratic telescope allocations are typical.  As a result, the statistical analysis of variable stars is relegated to a handful of measures: Welch-Stetson measure variability amplitude, skewness and kurtosis to evaluate non-Gaussianity of flux values, Edelson-Krolik discrete correlation function to study autocorrelation, sigma-clipping to locate outliers, and Gaussian Processes regression to smooth over gaps (Feigelson \& Babu 2012).   Periodic behaviors are sought from the phase dispersion minimization, Lomb-Scargle, Box Least Squares and Transit Comb Filter periodograms.  The mathematical foundations for statistics applied to irregularly spaced time series are often insecure and statistical probabilities for the variability measures are often unreliable (e.g. VanderPlas 2018).  

However, a vast methodology becomes available for sufficiently rich (hundreds or more points), regularly spaced time series.  They have been developed over the past half-century by professional methodologists (statisticians, engineers and economists) to study terrestrial and human-generated time domain phenomena, resting on solid mathematical foundations developed in hundreds of papers in dozens of journals.  The methods are described in texts such as Box et al (2015), Chatfield (2004), Enders (2014), Hyndman et al. (2014), and Shumway \& Stoffer (2017).   Methods like maximum likelihood ARIMA (autoregressive integrated moving average) have over a million Google hits, yet are rarely used in astronomy as they are designed for evenly spaced data.   Yet low-dimensional ARIMA models have proved to be very successful in modeling the micro-variability of ordinary stars using regular cadence lightcurves from NASA's Kepler mission (Caceres et al. 2019).  Hundreds of methods have reliable code implementations in the public domain R (R Core Team 2018) statistical software environment; each can be called with a simple line in a compact script.  

Most of the surveys planned for LSST, including the Deep Drilling Fields, will have sparse and unevenly spaced cadences.  They will not alleviate the weakness of methodology available for analyzing variable stars.  But LSST is capable of obtaining a remarkable dataset of $\sim 1$ million stellar lightcurves, each with $\sim 3500$ observations in a regular cadence.  This can be achieved by observing for a single low-Galactic latitude field for a single long winter night using a single photometric filter.  A powerful array of statistical analyses can then analyze this ensemble of well-cadenced lightcurves, giving an unprecedented view of short-timescale stellar variability, revealing both typical and unusual variability behaviors in an unbiased fashion.

The main limitation of this project is the 14-hour duration: only rapid variations can be characterized.  Many classes of variables with timescales of days, weeks or years will be poorly characterized and identified.  There is no way to mitigate this problem without building other LSSTs in Africa and Australia like the HAT-South project, or placing LSST in the South Pole like the AST3 project.  

Time series analysis of regularly cadenced data usually involves multistage procedures such as kernel or Gaussian Processes smoothing, autoregressive modeling, wavelet analysis with thresholded denoising, Fourier analysis with smoothing and tapering, as well as modern signal processing techniques like dynamic time warping, Hilbert-Huang transform and Singular Spectrum Analysis.  While these can be performed on the lightcurves obtained here, the very large sample precludes individualized study.   Therefore individual star analysis is supplemented by an ensemble approach involving extraction of scalar values from the lightcurves and analysis products, such as signal-to-noise ratio of Fourier and wavelet peaks.  These can be combined with a variety of time series diagnostics that are inherently scalar such as probabilities of statistical tests:  Anderson-Darling test for normality, Durbin-Watson test for serial autocorrelation, Ljung-Box portmanteau test for autocorrelation, augmented Dickey-Fuller and Kwiatkowski-Phillips-Schmidt-Shin tests for stationarity, Engle's ARCH test for heteroscedasticity, and Brock-Dechert-Scheinkman test for nonlinearity.  A suite of methods for `change point analysis' (Tartakovsky et al. 2014) can be added to search for sudden changes in temporal behaviors, as occurs in flaring and accretion stellar systems.  These are all coded in R's CRAN packages.

The product of such a multi-faceted analysis of the single-night lightcurves would be a multivariate dataset with $\sim 1$ million rows and dozens of `features'.  Many entries will be inaccurate or unreliable for the fainter stars around $R \simeq 22-24$.  But the ensemble will constitute the most comprehensive and unbiased characterization of rapid stellar variability ever obtained in astronomy.  Then methods of multivariate clustering can be applied to find $-$ objectively and without historical bias $-$ both rich and sparse classes of stellar variables.  These methods are mostly non-parametric including traditional approaches (such as hierarchical agglomerative clustering, $k$-means partitioning, Voronoi tessellations) and newer methods designed for Big Data (such as CLARA, CURE, BIRCH, and HDBSCAN).  These methods are described Everitt et al. (2011) and Xu \& Wunsch (2009) with R/CRAN software implementations.  Note that machine learning classification procedures like Random Forests cannot be used here, as there are few  training sets for known variable types with lightcurves similar to that obtained in the single-night LSST exposure.  

This single-night observation can be the basis for several immediate science papers, in addition to followup studies:  
\begin{description}

\item [Characterization of $\sim 1$M stellar lightcurves over 14 contiguous hours] ~~ This is a unique scientific dataset, never before achieved or envisioned, giving the most detailed and unbiased view of stellar variability on short timescales. Histograms and scatter plots of lightcurve features will be presented and interpreted.    Implications for the impact of stellar activity on planet habitability will be discussed.  An atlas of interesting lightcurves and a machine-readable table of scalar properties for $\sim 1$M stars will be provided in an archive site (e.g. IoP, MAST, Zenodo). 

\item [Objective identification of classes of rapidly variable stars] ~~ Variable stars are usually placed into historical categories (e.g. BY Dra, T Tau, Cepheid, cataclysmic, PG1159 variables) using heuristic methods.  A multivariate analysis of the features, using both nonparametric and parametric statistical methods of clustering, will give the first unbiased and quantitative identification of major and minor variability classes.  Later research can associate these data-derived classes with historical labels.  Unusual outliers will be identified.  A machine-readable table with classifications for $\sim 1$ million stars may give a large transformative boost to variable star studies. 

\item [Statistical study of late-type stellar flares]  ~~ Roughly half of the stars will be dM stars, the most common type in the Galaxy. Most will have ages $> 1$ Gyr when flaring activity is low, but many thousands will be younger and magnetically active.  Multimodality in flaring activity (e.g. due to the transition to fully convective interiors later than dM4) will be sought.  

\item [Identification of possible exoplanetary transits] ~~ While very few planets have sufficiently short orbital periods to show multiple transits in 14 hours, we expect that many stars will show single dips with a box-like shape characteristic of exoplanet transits.  A specialized analysis will be performed to identify and count these stars. 
\end{description}

But there may be another outcome from this study that may assist time domain studies based on the more common sparse, irregular cadence LSST survey lightcurves.  The clustering effort will have a dimensional reduction stage where some features are found to be extraneous and others are effective in discriminating classes.  The most useful features can then guide statistical characterization of the irregular lightcurves, speeding classification as the all-sky survey progresses.  These results may be of particular interest to LSST Event Brokers that will try to classify variable stars.  Finally, the R codes (with Python wrappers) for this project will be quickly released to assist other LSST time domain astronomers. 

In summary, this proposal uses a novel $-$ but very simple cadence $-$ to give a large, unique time domain dataset characterizing short-term stellar variability.  The ensemble of lightcurves will give:  unique insight into stellar variability issues such as the distribution of magnetic flares in main sequence stars; a new objective classification of rapid variability in stars;  and opportunity to discover rare new species with variations on rapid (15~sec to 14~hrs) timescales.  The machine learning classifier will be available for the full LSST community (e.g. LSST Event Brokers), improving classification for the dominant irregularly cadenced surveys.    Though a similar observation could be made with another telescope, data acquisition with LSST optics, camera, and filter set will greatly improve its applicability to other LSST programs.  Essentially, this one-night observation can provide the gold standard of LSST variability studies on short timescales. 
}

\clearpage

\section{Technical Description}

\subsection{High-level description}

An LSST field at low Galactic latitude will be observed continuously for a long winter night ($\sim 14$ hr) with $\sim 3500$ regularly spaced 15 sec exposures in a single photometric band.   An additional 30~min allocation is requested for a different night on the same field to obtain 5~min photometry in 6-bands to assist with science analysis.  
\vspace{.3in}

\subsection{Footprint -- pointings, regions and/or constraints}

The suggested winter sky location is  $(l,b) = (330^\circ, +20^\circ)$ chosen to give an appropriate stellar density with low absorption and a mixture of Pop I and Pop II stars.  This pointing should be revised based on LSST staff advice on star crowding (easily rectified by increasing the latitude) and saturation by bright stars.  Also the field location may be revised from scientific considerations after consultation with the full LSST Transient \& Variable Star (TVS) Science Collaboration.  

\subsection{Image quality}

According to a stellar population synthesis simulation from the Besan\c{c}on model \\ (http://model2016.obs-besancon.fr), a 9.6 deg$^2$ LSST field located at $(l,b) = (330^\circ, +20^\circ)$ with a sensitivity limit $r < 24.0$ magnitude will have $\sim 1.2$ million stars. 0.7M of them have $r<22$ to give high S/N lightcurves.  Most of these are at distances between 1 and 8 kpc, median absorption is $A_V \sim 0.2$ mag, and ages $3-10$ Gyr,   Most are K and M stars with $\sim 30,000$ F-G stars. 

The field location should be revised based on LSST staff advice regarding an appropriate stellar density to avoid overlapping PSFs given the expected image quality.  

\subsection{Individual image depth and/or sky brightness}

Image depth not critical.  Image quality must be sufficient to avoid common overlaps of stellar PSFs.  The field location (Galactic latitude) can be changed to reduce star density to a safe level.

\subsection{Co-added image depth and/or total number of visits}
One visit, no co-adds

\subsection{Number of visits within a night}
A continuous sequence of $\sim 3500$ 15 second exposures during a long winter night, from dusk until dawn.  

\subsection{Distribution of visits over time}
One all-night visit, plus a 30~min visit on a different night on the same field to obtain 5~min photometry in 6-bands to assist with science analysis.

\subsection{Filter choice}
$r$ band for the all-night visit.  $ugrizy$  bands for the 30-min allocation. 

\subsection{Exposure constraints}
A continuous sequence of $\sim 3500$ 15 second exposures during a long winter night, from dusk until dawn

\subsection{Other constraints}
None

\subsection{Estimated time requirement}
A single long winter night ($\sim$14 hours) plus 30 minutes on another night for 6-band photometry.

\vspace{.3in}

\begin{table}[ht]
    \centering
    \begin{tabular}{l|l|l|l}
        \toprule
        Properties & Importance \hspace{.3in} \\
        \midrule
        Image quality &  2   \\
        Sky brightness &  3\\
        Individual image depth &  2 \\
        Co-added image depth &  3 \\
        Number of exposures in a visit   &  3 \\
        Number of visits (in a night)  &  1 \\ 
        Total number of visits &  1 \\
        Time between visits (in a night) &  1\\
        Time between visits (between nights)  & 3  \\
        Long-term gaps between visits & 3\\
        Other  & Continuous 15 sec exposures  \\ 
         & in one night, R band \\
        \bottomrule
    \end{tabular}
    \caption{\bf Constraint Rankings:}
        \label{tab:obs_constraints}
\end{table}

\subsection{Technical trades}
\begin{enumerate}
    \item What is the effect of a trade-off between your requested survey footprint (area) and requested co-added depth or number of visits?  \\ {\it Not Applicable}
    \item If not requesting a specific timing of visits, what is the effect of a trade-off between the uniformity of observations and the frequency of observations in time? e.g. a `rolling cadence' increases the frequency of visits during a short time period at the cost of fewer visits the rest of the time, making the overall sampling less uniform.  \\ {\it Continuous cadence is essential}
    \item What is the effect of a trade-off on the exposure time and number of visits (e.g. increasing the individual image depth but decreasing the overall number of visits)?\\ {\it Continuous cadence is essential }
    \item What is the effect of a trade-off between uniformity in number of visits and co-added depth? Is there any benefit to real-time exposure time optimization to obtain nearly constant single-visit limiting depth? \\ {\it Not Applicable}
    \item Are there any other potential trade-offs to consider when attempting to balance this proposal with others which may have similar but slightly different requests? \\ {\it Not Applicable }
\end{enumerate}

\section{Performance Evaluation}

If LSST photometry operates as expected, no difficulties in performing the Special Data Processing and ensuing scientific analysis is expected.  The science will be compromised only if observational or instrumental conditions are unusually poor.  

The LSST operations staff will be consulted to see if the unusual program $-$ all-night observation without changes in pointing or photometric band $-$ poses any difficulties.  The science will not be strongly mitigated if the scheduling procedures do not give a strictly evenly-spaced cadence (e.g., if times between exposures fluctuate between 15 and 17 seconds rather than exactly 15 seconds).  The data will be forced onto a fixed time grid and small gaps can be imputed using established statistical techniques.

\vspace{.6in}

\section{Special Data Processing}

The required input data for the science goals are standard tables of calibrated R magnitudes for each of $\sim 3000$ 15 second exposures during a single night.  No special LSST data processing is needed.

Each regularly-spaced lightcurve will then be analyzed to acquire dozens of `features' characterizing its variability in different ways, and these features will then be subject to unsupervised clustering algorithms.  All of the required methods are already coded in the comprehensive, public domain R statistical software environment that has thousands of functions for analysis of regularly spaced time seres.  See the summary of $\sim 200$ R packages in the {\it CRAN Task View: Time Series Analysis} at https://cran.r-project.org/web/views/TimeSeries.html.   Then features will then be subject to nonparametric clustering algorithms.  Again these are already coded in R and its CRAN packages (see \\ https://cran.r-project.org/web/views/Cluster.html).  

The coding effort consists of scripting calls to CRAN packages, plotting, and data wrangling.  Penn State already has R scripts for time series analysis of astronomical time series and for clustering that can be adapted to this effort.  A reasonable estimate of labor is $\sim 2$ weeks for scripting and testing, and $\sim 1$ month wall-clock for data processing.  The scientific interpretation and communication of results will be longer, involving volunteer members of the LSST TVS (Transients \& Variable Stars) and ISSC (Information \& Statistics) Science Collaborations.  

A rough estimate of the computational load is 5 CPU-minutes/star, or $\sim 100,000$ CPU-hours for the full dataset using Intel Xeon E5-2680 processors.  (Probably many of the faintest $R \sim 23-24.5$ stars will not be fully analyzed, so the computation will be lighter.)  The problem is `embarrassingly parallel' so an arbitrary number of cores can operate independently.  The full output of intermediate results is likely to be $\sim 1$ TBy.  The computation will be performed, free of charge to LSST, on Penn State's Institute of CyberScience Advanced Cyber Infrastructure and CyberLAMP supercomputers.  

\section{References}
{\small
Box, G. E.~P., Jenkins, G.~M., Reinsel, G.~C., \& Ljung, G.~M. 2015, {\it Time Series Analysis: \\ \hspace*{0.3in} Forecasting and Control}, Wiley \\
Caceres, G. A., Feigelson, E. D., et al. 2019, AutoRegressive Planet Search: Methodology, \\ \hspace*{0.3in}  submitted \\
Chatfield, C. 2004, {\it The analysis of time series: an introduction}, 6th ed., CRC Press \\
Everitt, B. S., Landau, S., Leese, M.\ \& Stahl, D.\ 2011, {\it Cluster Analysis}, 5th ed., Wiley \\
Feigelson, E. D.\ \& Babu, G. J.\ 2012, {\it Modern Statistical Methods for Astronomy with R \\ \hspace*{0.3in} Applications}, Cambridge University Press\\
Hyndman, R.~J., \& Athanasopoulos, G. 2014, {\it Forecasting: Principles and Practice}, OTexts, \\ \hspace*{0.3in} https://www.otexts.org/fpp \\
Kuhn, M.~A., Hillenbrand, L.~A., et al. 2018, arXiv:1807.02115 \\
R Core Team 2018, R: A language and environment for statistical computing. R Foundation for Statistical Computing, Vienna, Austria. https://www.R-project.org\\
Shumway, R.~H.\ \& Stoffer, D.~S.\ 2017, {\it Time Series Analysis and Its Applications: With R \\ \hspace*{0.3in} Examples}, 4th ed.\\
Tartarkovsky, A., Nikiforov, I.\ \& Basseville, M.\ 2014, {\it Sequential Analysis: Hypothesis Testing \\ \hspace*{0.3in} and Changepoint Detection}, Chapman \& Hall \\
VanderPlas, J.~T.\ 2018, {\it ApJ Suppl}, 236, \#16 \\
Xu, R.\ \& Wunsch, D. C.\ 2009, {\it Clustering}, IEEE Press, Wiley
}
\end{document}